\documentclass[twocolumn,prb,showpacs,amsmath,amssymb,floatfix]{revtex4}
\usepackage{graphicx}
\usepackage{dcolumn}
\usepackage{bm}

\usepackage{makeidx}
\usepackage{epsfig}

\begin{document}

\title{Nonmonotonic energy harvesting efficiency in biased exciton chains}

\author{S. M. Vlaming}
\affiliation{Centre for Theoretical Physics and Zernike Institute
for Advanced Materials, University of Groningen, Nijenborgh 4, 9747
AG Groningen, The Netherlands}

\author{V.\ A.\ Malyshev}
\affiliation{Centre for Theoretical Physics and Zernike Institute
for Advanced Materials, University of Groningen, Nijenborgh 4, 9747
AG Groningen, The Netherlands}

\author{J.\ Knoester}
\affiliation{Centre for Theoretical Physics and Zernike Institute
for Advanced Materials, University of Groningen, Nijenborgh 4, 9747
AG Groningen, The Netherlands}

\begin{abstract}
We theoretically study the efficiency of energy harvesting in linear
exciton chains with an energy bias, where the initial excitation is
taking place at the high-energy end of the chain and the energy is
harvested (trapped) at the other end. The efficiency is
characterized by means of the average time for the exciton to be
trapped after the initial excitation. The exciton transport is
treated as the intraband energy relaxation over the states obtained
by numerically diagonalizing the Frenkel Hamiltonian that
corresponds to the biased chain. The relevant intraband scattering
rates are obtained from a linear exciton-phonon interaction.
Numerical solution of the Pauli master equation that describes the
relaxation and trapping processes, reveals a complicated interplay
of factors that determine the overall harvesting efficiency.
Specifically, if the trapping step is slower than or comparable to
the intraband relaxation, this efficiency shows a nonmonotonic
dependence on the bias: it first increases when introducing a bias,
reaches a maximum at an optimal bias value, and then decreases again
because of dynamic (Bloch) localization of the exciton states.
Effects of on-site (diagonal) disorder, leading to Anderson
localization, are addressed as well.

\end{abstract}


\maketitle

\section{Introduction}
\label{Sec: Introduction}

The harvesting of electromagnetic energy, i.e., its absorption,
transport to a specific site, and subsequent trapping in an
alternative form of energy, is a process of great importance to life
on earth. In nature, systems occur that perform these tasks with
amazing efficiency. In order to make the entire process
unidirectional, a certain downward energy gradient (bias) occurs. An
excellent example is the light-harvesting system in purple bacteria,
where photons with a wavelength around 800 nm may be absorbed in the
B800 ring of the light-harvesting system 2 (LH2). The resulting
excitation may be transferred to the lower-energy B850 ring (850
nm), before undergoing transfer to the yet lower-energy LH1 system
(875 nm), from where the final transfer to the so-called special
pair of the reaction center occurs ("the trap"). Here, the
excitation energy is transformed into chemical energy (see
Refs.~\onlinecite{Amerongen00} and \onlinecite{Sundstrom99} for a
review). All these steps occur rapidly and with extremely high
efficiency. It is well-understood by now that the transport in this
system is not described by the classical F\"orster
mechanism~\cite{Foerster48,Agranovich82} of incoherent energy
transfer. The excitonic nature plays an important role, both for
transport inside and between the various
subsystems.~\cite{Sundstrom99,Oijen99,Sumi99,Mukai99,Silbey06}

Also in man-made nanoscale systems, the injection of energy at some
point and its subsequent transport to and collection at another
site, is a process of growing interest. Not only is this relevant to
artificial light-harvesting systems, but also to electronic and
photonic nanodevices. Obviously, it is important to study how the
efficiency of the total process can be optimized.

The aim of the present paper is to study theoretically the effect
of a downward energy gradient (bias) on the efficiency of energy
harvesting in linear chains of strongly coupled two-level units,
and thus to investigate how this process may be optimized. In the
context of this paper, we consider harvesting as the total process
following initial excitation at the high-energy side of the chain
until its trapping at the low-energy side, i.e., it includes both
the transport and the trapping processes. We will choose the
average trapping time as a measure for the efficiency of the
harvesting process.

At first glance, it seems obvious that a downward bias will reduce
the harvesting timescale, and thus will increase its efficiency.
Indeed, in the F\"orster limit of incoherent hopping between weakly
coupled sites, an energy bias will cause the excitation to diffuse
towards the lower energy end of the chain and will generally lead to
a higher quantum efficiency for the harvesting process. Experimental
realizations of such chains have been studied by Sauer and
coworkers;\cite{Heilemann04,Tinnefeld05} in addition low-generation
dendritic systems with an energy bias towards their core have been
studied by various groups.\cite{Devadoss96,Jiang97,Kopelman97}

In contrast to the weak-coupling case, the effect of a bias is not
clear a priori in chains with strong intersite coupling, where in
the absence of a bias the excited states are delocalized (excitons)
over the entire chain. On the one hand, the bias defines a
preference for the excitation to move towards the trap, on the other
hand, it leads to dynamic (Bloch) localization of the exciton
states,~\cite{Bloch28,Zener34} which slows down the transport
process. We will show that, as a result of the competition between
both effects, a strong bias always decelerates the harvesting
process, while for smaller values of the bias, nonmonotonic behavior
may arise. We also study the effect  of energetic disorder, which
causes Anderson localization\cite{Anderson58,Abrahams79} and thereby
also acts toward reducing the transport efficiency.

To model the harvesting process, we will describe the system by a
Frenkel exciton Hamiltonian, with site energies that follow a linear
energy bias and, superimposed on that, may suffer from random
disorder. The initial state is a localized excitation of a single
monomer at the high-energy end of the chain, while at each time the
rate of trapping is proportional to the probability that the monomer
at the low-energy end of the chain (the trap) is excited. Thus, to
get trapped the exciton should either overlap appreciably with the
trap or travel over the chain until it does so. The transport is
modeled as an intraband relaxation process, induced by the
scattering of the exciton eigenstates on phonons, and described by a
Pauli master equation for the exciton populations. We will restrict
ourselves to the zero-temperature limit. A study of temperature
effects will be presented elsewhere.\cite{Vlaming07}

Possible realizations of our model are molecular
aggregates~\cite{Kobayashi96,Knoester02} and conjugated
polymers~\cite{Tilgner92,Hadzii99} subjected to a non-uniform
electric field in order to create a bias (Stark effect), molecular
photonic wires~\cite{Wagner94,Heilemann04,Tinnefeld05} in which a
bias is created by chemical synthesis of a series of molecules with
different transition energies, linear arrays of metal
nanoparticles,\cite{Quinten98,Citrin04} and assemblies of
semiconductor quantum dots.~\cite{Crooker02,Franzl04,Klar05} In the
latter two examples, a bias may be created by carefully arranging
particles of different size. Another possible application of the
underlying physics might be the electrical transport through a
single DNA molecule;~\cite{Porath00,Xu04,Gutierrez05,Malyshev07} the
source-drain voltage applied to the DNA is usually on the order of a
few volts, which is high enough for dynamic
localization to occur. %[keep this DNA here? Shift to
%conclusions?]

The outline of this paper is as follows. In the next section, we
present our model Hamiltonian. The Pauli master equation for the
population dynamics as well as the trapping model are introduced in
Sec.~\ref{Sec: Population dynamics}. In Sec.~\ref{Sec: Numerics}, we
present our numerical results, analyzing first the effect of a bias
on the exciton trapping time in disorder-free chains
(Sec.~\ref{Subsec: Homogeneous}). The interplay of diagonal disorder
and a bias is unraveled in Sec.~\ref{Subsec: Inhomogeneous}. We
summarize and make final remarks in Sec.~\ref{Sec: Summary}.

\begin{figure}[ht]
    \centerline{\includegraphics[width=8cm,clip]{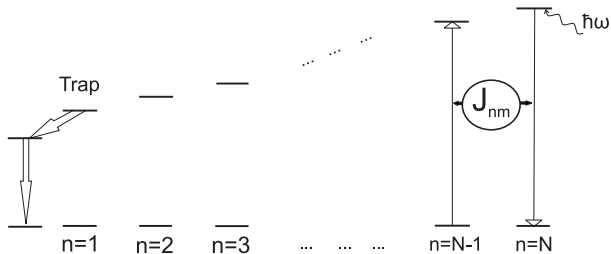}}
    \caption{Schematics of the system under consideration: a
    one-dimensional array of two-level units with a linearly
    varying excitation energy. Static disorder can be present
    as well, introducing an additional stochastic variation in the
    excitation energies. The initial excitation is created at the
    high-energy site $n=N$. The excited state of site $n=1$ is
    strongly quenched by an additional molecule, so that it acts
    as a trap.}
    \label{fig:schema}
\end{figure}

\section{Model Hamiltonian}
    \label{Sec: Exciton Hamiltonian}

We consider a one-dimensional chain of $N$ two-level units with
aligned transition dipoles, which are coupled through
dipole-dipole transfer interactions. The site excitation energies,
in addition to a stochastic component (diagonal disorder), include
a linearly varying part (bias). The schematics of the system is
depicted in Fig.~\ref{fig:schema}. The low-energy end of the chain
contains a trap, while energy is injected at the high-energy end.
In Sec.~\ref{Sec: Population dynamics}, we will define how the
transport and trapping processes take place. In this section, we
consider the relevant Hamiltonian, which reads
\begin{equation}
    \label{frenkelham}
    H_{ex}=\sum_{n=1}^N [\varepsilon_n + (n-1)\Delta]\left|n\right>\left<n\right|
    + \sum_{n,m\ne n}^N J_{nm} \left|n\right>\left<m\right|,
\end{equation}
Here, the state $\left|n\right>$ corresponds to site $n$ being
excited, while all other sites are in the ground state. The site
excitation energy contains two parts: (i) $\varepsilon_n$, which is
a Gaussian stochastic variable with mean $\bar{\varepsilon}$ (which
can be set to zero without loss of generality) and standard
deviation $\sigma$, and (ii) the energy bias, characterized by the
parameter $\Delta$. The transfer integrals $J_{nm}$ are given by
$J_{nm}=-J/|n-m|^3$, with the nearest-neighbor coupling $-J < 0$.

The exciton eigenenergies $E_s~(s=1,...,N)$ and eigenfunctions
$\left|s\right>=\sum_{n=1}^N c_{sn}\left|n\right>$, with $c_{sn}$
being the $n$th component of the $s$th eigenfunction, are found by
diagonalizing the matrix $H_{nm}=\left< n\right|H\left|m\right> =
\delta_{nm}[\varepsilon_n + (n-1)\Delta] + J_{nm}$ and can be taken
real in our model. In the disorder- and bias-free case ($\sigma =
\Delta = 0$), the exciton states are delocalized over the whole
chain. As a convenient reference case, we note that the wave
functions $|s\rangle$ resemble those obtained when only the
nearest-neighbor coupling is taken into
account:~\cite{Fidder91,Malyshev95,Didraga04}
\begin{equation}
    \label{golffunctie, geen bias}
    |s \rangle = \left(\frac{2}{N+1}\right)^{1/2}\sum_{n=1}^N
    \sin\left(\frac{\pi s n}{N+1}\right) |n \rangle \ .
\end{equation}
The $N$ exciton states form a band with approximate energies
\begin{equation}
    \label{excenergy}
    E_s = -2J \sum_{n=1}^N \frac{1}{n^3}
    \cos\left(\frac{\pi s n}{N+1}\right) \ ,
\end{equation}
that range from $E_1 = -2.404 J$ to $E_N = 1.803 J$, thus having a
width $E_N - E_1 = 4.207 J$.~\cite{Fidder91} Note that $E_s$
increases with $s$ ($J>0$). Throughout this paper (i.e., also in the
presence of bias and/or disorder), we will use the convention that
$s$ labels the states in order of growing energy. Although
Eqs.~(\ref{golffunctie, geen bias}) and (\ref{excenergy}) are useful
for reference, it should be stressed that all results in this paper
are obtained by exact numerical diagonalization, accounting for all
dipole-dipole interactions.

In a disorder-free system, a nonzero bias results (in the
thermodynamic limit $N \to \infty$) in dynamic (Bloch) localization
of all states.~\cite{Bloch28,Zener34} The localization size is
estimated from semiclassical arguments as $L_B = (E_N -
E_1)/\Delta$.~\cite{Ashcroft76} The Bloch localization is
accompanied by the subsequent reorganization of the energy spectrum
of the system, which becomes ladder-like with level spacing
$\Delta$.~\cite{Wannier60} This structure is revealed in
photoluminescence~\cite{Mendez88,Agullo89} and
photoconductivity~\cite{Saker91} spectra of semiconductor
superlattices as a series of equally spaced peaks.

\begin{figure}[ht]
    \centerline{\includegraphics[width=8cm,clip]{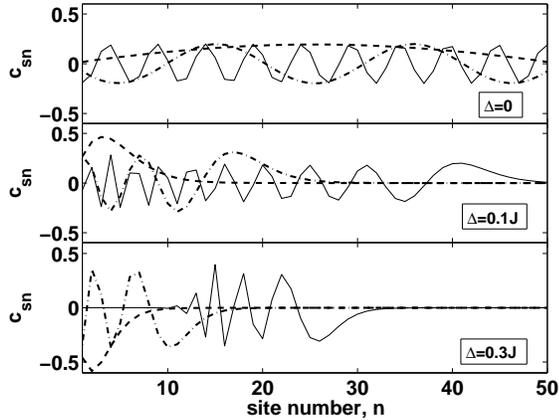}}
    \caption{Exciton wave functions of a disorder-free chain of
    size $N = 50$ for the states $s = 1$ (dashed curves), $s = 5$
    (dash-dotted curves), and $s = 25$ (solid curves) calculated by
    diagonalizing the Hamiltonian~(\protect\ref{frenkelham}) for various
    magnitudes of the bias: $\Delta = 0$ (upper panel), $\Delta = 0.1 J$
    (middle panel), and $\Delta = 0.3 J$ (lower panel). The lines
    are a guide for the eye; the coefficients $c_{sn}$ are
    only defined for integer values of $n$.}
    \label{fig:f2}
\end{figure}

In finite-size systems, dynamic localization comes into play when
$L_B$ becomes smaller than the system size $N$ or, in other words,
when the overall bias $N\Delta$ exceeds the bare exciton bandwidth
$4.207 J$. Figure~\ref{fig:f2}, where we plotted several wave
functions calculated for three magnitudes of the bias, illustrates
this. Note that for $L_B < N$ there exists a strong correlation
between the energy of an exciton and its location: the states with
smaller $s$ (near the bottom of the exciton band) are localized at
the lower energy end of the chain, while the states with higher $s$
(near the top of the band) are localized at the higher energy end.
Upon increasing $\Delta$, the bandwidth approaches a value
$N\Delta$~.~\cite{Diaz06}

Introducing a stochastic component in the site energies
$\varepsilon_n$ will result in an additional (Anderson) localization
effect. At sufficiently high disorder strength $\sigma$, the
bias-induced correlations between the energy of a state and its
location on the chain will be destroyed. The lowest-energy exciton
states will not necessarily be located at the low-$n$ side of the
chain.

\section{Transport and trapping models}
    \label{Sec: Population dynamics}

\subsection{Formalism}
\label{Subsec: Formalism}

The proper description of the dynamics of quantum particles in a
system that interacts with a bath, requires the density matrix
formalism.~\cite{Neumann27,Blum81,May00} This holds in particular
for the transport and trapping processes underlying the energy
harvesting considered here. The system studied is described by the
reduced density matrix $\rho_{ss'}$ of the exciton system, which is
obtained from the total density operator by tracing out the bath. In
our case, the bath consists of the vibrations in the host
surrounding the exciton chain. The diagonal elements $\rho_{ss}$
represent the exciton populations, i.e., the probability that
exciton eigenstate $s$ is excited; the $\rho_{ss'}$ for $s \neq s'$
describe the coherences between different exciton states.

For sufficiently weak coupling between the excitons and the host
vibrations, we can use the Born-Markov approximation, where it is
assumed that the interaction with the host can be treated as a
perturbation and that the host is always in equilibrium. A further
simplification occurs if we only consider time scales that are long
compared to the typical oscillation times of the density matrix
elements. This is referred to as the secular
approximation,~\cite{Blum81} and leads to a decoupling of the time
evolution of the populations and the coherences.

The time evolution of the populations is then given by the Pauli
master equation and does not involve any coherences:
\begin{equation}
\label{pauli}
    \dot{\rho}_{ss} =\sum_{r \neq s}\rho_{rr}W_{sr}-\rho_{ss}
    \sum_{r \neq s} W_{rs} \ ,
\end{equation}
where the vibration-induced scattering
rates $W_{ss'}$ will be discussed later on. In the secular
approximation, the coherences $\rho_{ss'}$ decay exponentially with
the dephasing rate $\gamma_{ss'}= \frac{1}{2}\sum_r(W_{rs}+W_{rs'})$
while oscillating in time at a rate $\omega_{ss'}=E_s-E_{s'}$.
Additional terms will be included to account for the trap and
spontaneous decay and, in fact, the trapping term does introduce a
new coupling of the populations to the coherences, as will be seen
shortly. However, in most cases we consider, the dephasing of the
coherences is sufficiently rapid compared to the harvesting time to
allow for a treatment that involves the populations only. A further
discussion on the validity of this approach is presented in
Sec.~\ref{Subsec: Validity}.

As we already mentioned in the Introduction, the energy harvesting
process not only includes the transport, but also the trapping at
the low-energy end of the chain. We will assume that the trapping
does not affect the exciton wave functions and that this process
changes both the exciton populations and coherences with a rate that
is proportional to the probability that site $n=1$ is excited:
$\dot{\rho}_{nn'}^{(trap)}=-\frac{1}{2} \Gamma
(\delta_{n1}+\delta_{n'1})\rho_{nn'}^{(trap)}.$ We will refer to the
constant $\Gamma$ as the quenching or trapping amplitude. By
transforming $\dot{\rho}_{nn'}^{(trap)}$ to the exciton basis, it is
found that indeed both the exciton populations and coherences are
affected by the trap and that the action of the trap couples
populations to coherences. If on the timescale of the harvesting the
coherences may indeed be neglected, the population of any exciton
state $s$ decays nonradiatively with the rate
\begin{equation}
    \label{quench1}
    \Gamma_s = \Gamma c_{s1}^2 \ .
\end{equation}
After accounting for spontaneous emission by introducing a radiative
decay channel with rate $\gamma_s = \gamma_0 (\sum_n c_{sn})^2$
($\gamma_0$ is the radiative decay rate of an isolated monomer), the
overall exciton population dynamics can be described by the
following Pauli master equation for the exciton populations $P_s
\equiv \rho_{ss}$ :
\begin{equation}
\label{PME}
    \dot{P}_s = -(\gamma_s + \Gamma_s)P_s
    + \sum_{s'}\left( W_{ss^{\prime}}P_{s^{\prime}}
    - W_{s^{\prime}s}P_s \right) \ .
\end{equation}

As the initial condition to Eq.~({\ref{PME}), we will assume that
the rightmost site $n = N$ is excited at $t = 0$:
$\rho_{nn'}(0)=\delta_{nn'}\delta_{nN}$. A transformation to the
exciton basis yields $\rho_{ss'}(0)=c_{sN}c_{s'N}$. This corresponds
to a normalized initial exciton population distribution given by
$P_s(0)=c_{sN}^2$. Furthermore, our initial condition implies that
off-diagonal density matrix elements $\rho_{ss'}$ (exciton
coherences) are also created; the effect of these coherences on the
energy transport may be relevant when their dephasing times are
larger than or comparable in magnitude to the typical harvesting
times. This is the case for a sufficiently strong quenching
amplitude $\Gamma$ in combination with strongly delocalized exciton
states. We come back to this point in Sec.~\ref{Subsec: Validity}.

\subsection{Exciton scattering rates}
\label{Subsec: scattering}

In the Born-Markov approximation, the scattering rates $W_{ss'}$
which feature in Eq.~(\ref{pauli}) are equivalent to those obtained
using the Fermi golden rule, with the exciton-vibration interaction
serving as the perturbation. As stated above, we think of vibrations
in the chain's host medium as the ones mainly responsible for the
scattering. We will restrict ourselves to one-phonon processes and
use a scattering rate from state $s^{\prime}$ to state $s$ of the
form~\cite{Leegwater97,Shimizu01,Bednarz02}
\begin{eqnarray}
    W_{ss^{\prime}}  =  W_0\ S(\left|E_s - E_{s^{\prime}}\right|)
    \,\sum_{n=1}^N c_{s n}^2 c_{s^{\prime} n}^2 \,
    \nonumber\\
    \times  \left\{
\begin{array}{lr}
    n(E_s - E_{s^{\prime}}), &\quad E_s > E_{s^{\prime}}\ ,
\nonumber\\
\nonumber\\
    1+n(E_{s^{\prime}} - E_s), &\quad E_s < E_{s^{\prime}} \ 
\end{array}
    \right.
\label{1Wkk'}
\end{eqnarray}
%
%\begin{multline}
%    W_{ss^{\prime}} = W_0\ S(\left|E_s - E_{s^{\prime}}\right|)
%    \,\sum_{n=1}^N c_{s n}^2 c_{s^{\prime} n}^2 \,\\
%    \times
%    \begin{cases}
%    n(E_s - E_{s^{\prime}}), &\quad E_s > E_{s^{\prime}}
%\\
%\\
%    1+n(E_{s^{\prime}} - E_s), &\quad E_s < E_{s^{\prime}}
%\end{cases}\ .
%\label{1Wkk'}
%\end{multline}
%
Here, the prefactor $W_0$ is a measure of exciton scattering. The
spectral density $S(|E_s - E_{s^{\prime}}|)$ depends on the
details of the exciton-vibration coupling as well as on the
vibrational density of states. In particular, within the Debye
model for the vibrational modes, this factor is given by $S(E_s -
E_{s^{\prime}}) = (|E_s - E_{s^{\prime}}|/J)^3$ (see, e.g.,
Ref~\onlinecite{Bednarz02}). We use a Debye-like spectral density
with an exponential cut-off factor, $S(E_s - E_{s^{\prime}})=|(E_s
- E_{s^{\prime}})/J|^3 \exp(-|E_s - E_{s^{\prime }}|/\omega_c)$,
where $\omega_c$ is a cut-off frequency. Similar spectral
densities have been successfully used to fit the optical dynamics
in photosynthetic antenna
complexes.~\cite{Kuhn97,May00,Renger01,Brueggemann04} Note,
however, that the results that we will present later on only
weakly depend on the particular form of $S(E_s - E_{s^{\prime}})$.

The term $\sum_{n=1}^N c_{s n}^2 c_{s^{\prime} n}^2$ in
Eq.~(\ref{1Wkk'}) represents the overlap integral of exciton
probabilities for the states $s$ and $s^{\prime}$. It depends on
both the magnitude of the bias, $\Delta$, and the strength of the
disorder, $\sigma$. Because of the overlap integral, only hops
between neighboring localized states are efficient. This strongly
limits the exciton transport along the chain.

Finally, $n(E) = [\exp(E/T) - 1]^{-1}$ is the occupation number of
the vibrational mode with energy $E$ (the Boltzmann constant is
set to unity). Due to the presence of the factors $n(E)$ and
$1+n(E)$, the rates $W_{ss^{\prime}}$ meet the principle of
detailed balance: $W_{ss^{\prime}} = W_{s^{\prime}s}\exp[(E_s -
E_{s^{\prime}})/T]$. Thus, in the absence of decay channels, the
eventual exciton distribution is the Boltzmann equilibrium
distribution.

\subsection{Harvesting time}
\label{Subsec: tau}

The object of our study will be the harvesting time $\tau$ which we
define as follows. We first introduce the overall survival time of
an exciton as the standard expectation value:
\begin{equation}
\label{surv}
    \bar{\tau} = \int_0^{\infty}dt \; t \; \Big\langle
     - \sum_s \dot{P}_s(t) \Big\rangle
    = \int_0^{\infty}dt \Big\langle \sum_s P_s(t) \Big\rangle \ ,
\end{equation}
where the angular brackets denote the average over disorder
realizations. This quantity should be calculated for two
situations: with ($\bar{\tau}$) and without ($\tau_0$) the trap.
Finally, the exciton harvesting time $\tau$ is obtained by
extracting the survival time $\tau_0$ with respect to the
radiative decay from the overall survival time $\bar{\tau}$,
according to the rule:~\cite{Malyshev03}
\begin{equation}
\label{taucorrect}
    \frac{1}{\tau} = \frac{1}{\bar{\tau}} - \frac{1}{\tau_0} \ .
\end{equation}

In fact, calculating $\tau$ does not require a full solution of
the kinetic equation~(\ref{PME}). Indeed, its solution reads
\begin{equation}
\label{formalpaulisol}
    P_s(t) = \sum_{s'}(e^{-\hat{R}t})_{ss'}P_{s'}(0) \ ,
\end{equation}
where we have introduced a matrix $\hat{R}$ with elements
\begin{equation}
\label{Rmatrix}
    R_{ss'}=(\gamma_s + \Gamma_s + \sum_{s''}W_{s''s})\delta_{ss'}-W_{ss'} \ .
\end{equation}
Using Eq.~(\ref{formalpaulisol}), the integration over time in
Eq.~(\ref{surv}) can be performed formally, yielding
\begin{equation}
\label{tauR}
    \bar{\tau} = \Big\langle \sum_{s,s'}R_{ss'}^{-1}P_{s'}(0)\Big\rangle \ .
\end{equation}
Thus, to calculate $\tau$ for a particular choice of parameters,
one has to invert two $N\times N$ matrices $\hat{R}$, namely one
for the chain with trap and one excluding the trap.

\subsection{Validity of the formalism}
\label{Subsec: Validity}

Throughout most of this paper, only the exciton populations are
considered, which in most cases is an excellent approximation. The
limit $\Gamma \gg W_0$, however, highlights some limitations of our
model, which are caused by not taking the coherences $\rho_{ss'}$
into account. In most situations we consider, this is justified, as
the coherences $\rho_{ss'}$ decay with a dephasing rate of an order
of magnitude of $\sum_{r}(W_{rs}+W_{rs'})$. This implies that as
soon as some relaxation process is involved in reaching the trap,
the harvesting process will be slow enough to allow for the neglect
of coherences. However, when the quenching amplitude $\Gamma$ is
large and the exciton states are delocalized over the entire chain
(i.e., short chains, small bias and small disorder), our model also
produces a clearly unphysical artifact, namely the possibility of
instantaneous energy transfer from the site $n=N$ to the trap. This
can be clearly seen from the quenching term in the Pauli master
equation (\ref{PME}): $-\Gamma_sP_s$ is non-zero at $t=0$ for the
initially populated exciton states. This problem is remedied by
including coherences in the trapping term, which in turn leads to
coupling between different coherences and populations.

We can estimate when the coherences may safely be neglected by
requiring that the probability of the initial population
distribution undergoing relaxation should be larger than the
probability of being directly caught by the trap: $\sum_s
\left(\sum_{s'}W_{s's}c_{sN}^2\right) \gg \sum_s \Gamma c_{s1}^2
c_{sN}^2$. For the case of no bias and no disorder, we can use the
(approximate) explicit expressions for $c_{sn}$ in Eq.
(\ref{golffunctie, geen bias}) to numerically estimate the above
inequality. The left hand side, which depends on the form of the
scattering rate, numerically evaluates to a $N$-independent value of
approximately $W_0/20$, while the right hand side can be estimated
as $\frac{3\Gamma}{2N}$. This estimate shows that our formalism is
valid for $\Gamma < NW_0/30$, which for the chain length $N=50$ and
the scattering amplitude $W_0=10J$ used later on translates into
$\Gamma < 17J$. Note that at larger biases Bloch localization
strengthens the validity of our model, as the delocalized exciton
states extending from the initially excited monomer to the trap that
are required for direct quenching (i.e., with no relaxation
involved) become increasingly rare and relaxation will thus be
necessary. In that case, the coherences will dephase sufficiently
fast to become irrelevant on the total timescale of the harvesting.

In order to get a more quantitative impression of the effect of
coherences, we have performed some additional calculations in a
crude model where, within the secular limit, coherent effects on the
trapping were taken into account by assuming a simple exponential
decay of the coherences, while both population and coherence
trapping terms were included:
\begin{subequations}
\begin{equation}
    \dot{\rho}_{ss}^{(trap)}=-\Gamma c_{s1}^2
    \rho_{ss}^{(trap)}-\frac{1}{2} \Gamma c_{s1}\sum_{r \neq s} c_{s'1}
    \left(\rho_{ss'}+\rho_{s's}\right) \ ,
\end{equation}
\begin{eqnarray}
    \dot{\rho}_{ss'}= & - & i \omega_{ss'}\rho_{ss'} - \frac{1}{2}
    \sum_{r}(W_{rs}+W_{rs'})\rho_{ss'} \nonumber\\ \nonumber\\  & - &
    \frac{1}{2}\Gamma \left(c_{s1}^2+c_{s'1}^2\right)\rho_{ss'}\ .
\end{eqnarray}
\end{subequations}
These equations are obtained by disregarding all terms that couple
the time evolution of the coherences to other density matrix
elements, while the equation for the time evolution of the
populations is complete and naturally also includes the phonon
scattering and radiative decay terms discussed previously. Although
this is not a consistent approach, it does allow for a first
impression of the effect of coherences on population decay. As we
will see in Sec.~\ref{Sec: Numerics}, the results of this approach
confirm the above estimates.

\begin{figure}[ht]
\centerline{\includegraphics[width=8cm,clip]{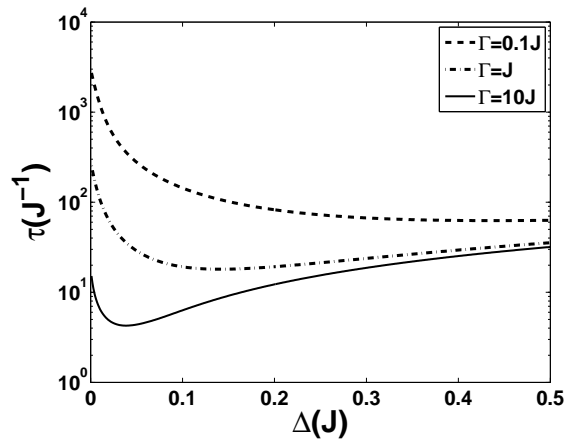}}
    \caption{Harvesting time $\tau$ [defined in Eq.~(\protect\ref{taucorrect})]
    versus the applied bias magnitude $\Delta$ calculated for a
    disorder-free chain of 50 sites for various quenching amplitudes $\Gamma$
    (shown in the plot). The exciton scattering
    amplitude was set to $W_0 = 10 J$. } \label{fig:quench2}
\end{figure}

\section{Numerical results and Discussion}
    \label{Sec: Numerics}

All results reported in this paper were obtained by using physical
parameters ($J, \gamma_0$, and $W_0$) typical for a particular
physical system where our theory might be applicable, namely
molecular aggregates. More specifically, we set $J =
600~\textrm{cm}^{-1}$ and $\gamma_0 = 3.6\times 10^8
\textrm{s}^{-1} = 2\times 10^{-5} J$, which are appropriate values
for J-aggregates of the dye pseudoisocyanine
(PIC).~\cite{Fidder90,Renge97} The scattering amplitude $W_0$ was
fixed at $W_0 = 10 J$, which is consistent with the values found
from the fit of the temperature dependence of the J-bandwidth and
of the radiative lifetime in J-aggregates of PIC.~\cite{Heijs05}
The cut-off frequency $\omega_c$ of the spectral density $S(E)$
was taken equal to $\omega_c = 0.5 J$. The exact value is
arbitrary to a large extent and hardly affects the conclusions
derived from our study, although it is crucial that some cut-off
value is introduced to suppress unphysically energetic phonon
modes. To simplify the link to other systems, the nearest-neighbor
coupling $J$ was used as the unit for all relevant quantities. We
stress once again that all dipole-dipole couplings are taken into
account. Chains of size $N = 50$ were considered in all
simulations. As was already mentioned in the Introduction, we only
consider the zero-temperature limit.

\subsection{Disorder-free biased chain}
    \label{Subsec: Homogeneous}

Figure~\ref{fig:quench2} shows the behavior of the harvesting time
$\tau$ as a function of the bias magnitude $\Delta$, calculated for
a set of quenching amplitudes $\Gamma$, indicated in the plot. As is
seen, the character of the behavior is strongly affected by
$\Gamma$, revealing an acceleration of the harvesting process with
increasing bias for $\Gamma < W_0 (= 10 J)$, while the
$\Delta$-dependence at $\Gamma \sim W_0$ is shown to be
nonmonotonic. For $\Gamma \gg W_0$, the model predicts a harvesting
time that monotonically increases with $\Delta$, starting from $\tau
\ll 1/J$ for $\Delta=0$. We do not show and analyze this limit
however, as it describes an unphysically strong quenching effect,
with $\Gamma$ being much larger than the nearest-neighbor coupling
$J$. Such strong quenching would affect the exciton wave functions;
moreover, the neglect of coherences is not a valid approximation in
this case (cf. Sec.~\ref{Subsec: Validity}).

To unravel the behavior found in Fig.~\ref{fig:quench2}, we note
that various bias-dependent factors exist that govern the overall
energy harvesting process: (i) - the initial distribution of
population $P_s(0) = c_{sN}^2$, (ii) - the trapping of the exciton
states with rate $\Gamma_s = \Gamma c_{s1}^2$, and (iii) - the
intra-band relaxation which populates (or depopulates) the most
strongly trapped states. The interplay of all of these factors
determines the bias dependence of the exciton harvesting presented
in Fig.~\ref{fig:quench2}.

We first discuss the initial condition, $P_s(0) = c_{sN}^2$, which
in Fig.~\ref{fig:f3}(a) is plotted as a function of the state index
$s$ for four choices of the bias strength $\Delta$. In the absence
of a bias ($\Delta = 0$), the states in the center of the band are
populated to a much larger extent than the states near the bottom
and the top of it. Similar behavior occurs in the nearest-neighbor
coupling model, where this can be explicitly shown by replacing
$c_{sN}$ with the $N$-th coefficient in Eq.~(\ref{golffunctie, geen
bias}). A nonzero bias drastically changes this situation, because
of the dynamic localization. The localization dictates that the
states which are located near the right-most site $n = N$, i.e.
those which have a significant amplitude at the site of initial
excitation, are those with higher energies or large $s$ (see
Fig.~\ref{fig:f2}). Thus, upon increasing the bias strength, the
initial distribution of population will shift to the top of the
exciton band. Figure~\ref{fig:f3}(a) clearly illustrates this.

\begin{figure}[ht]
\centerline{\includegraphics[width=8cm]{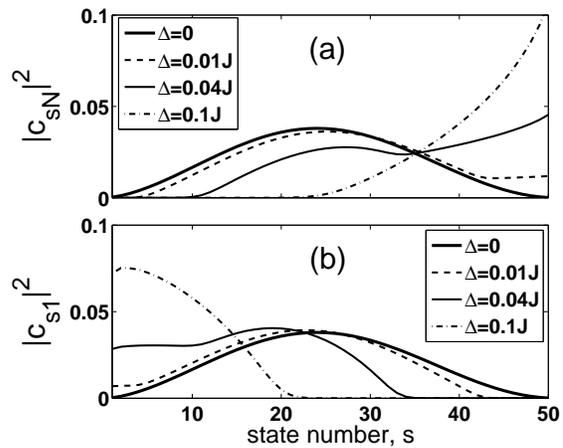}}
    \caption{Bias dependence of the exciton probabilities $c_{sN}^2$
    (a) and $c_{s1}^2$ (b) that determine the initial distribution of the
    exciton population $P_s(0)$ and the quenching rate $\Gamma_s$,
    respectively. The bias tends to shift the initial distribution
    of population to the top of the band (to higher $s$). Oppositely,
    the strongly quenched states are shifted to the lower band edge (small
    $s$) .}
\label{fig:f3}
\end{figure}

The distribution $c_{s1}^2$, which determines the quenching rate
$\Gamma_s$, shows the opposite tendency, as is seen in
Fig.~\ref{fig:f3}(b). In the absence of a bias, it is maximal for
the central band states and negligible at the band edges, similar to
$c_{sN}^2$. Already at a moderate bias, the quenching rate of the
states at the bottom of the band increases significantly. For larger
bias magnitudes, the distribution $c_{s1}^2$ strongly shifts to the
lower band edge (to small $s$). For a sufficiently strong bias
($N\Delta > E_N - E_1 = 4.207 J$), when the Bloch localization sets
in, the initially excited states and those undergoing efficient
quenching are separated completely and located at the top and the
bottom of the band, respectively.

Finally, an important stage in the harvesting process is the
relaxation of excitons from the initially populated states to those
where the quenching rates $\Gamma_s = \Gamma c_{s1}^2$ become
comparable to or larger than the relaxation rate $\sum_{s'}W_{s's}$.
This relaxation process will generally take longer when increasing
the bias magnitude $\Delta$. The reason for this is twofold. First,
as we have seen above, for larger bias the excitation and quenching
occur at opposite band edges, implying that the excitons have to
undergo more relaxation steps to be trapped. Second, the
bias-induced localization of the exciton states decreases their
probability overlap $\sum_n c_{sn}^2 c_{s'n}^2$, thereby reducing
the relaxation rates $W_{ss'}$.

The above observations form the basis for understanding the results
in Fig.~\ref{fig:quench2}. For small quenching amplitudes, $\Gamma
\ll W_0$, the downward relaxation is the fastest process and the
trapping step becomes rate-limiting. Eventually, most of the
population will end up in the exciton states near the lower band
edge, where it waits to be trapped. In the absence of a bias
($\Delta = 0$), these states have negligible quenching rates
$\Gamma_s$, as explained above, leading to large $\tau$ values.
After the bias is turned on, the amplitudes of the band edge states
at the trapping site $n = 1$ steeply increase, giving rise to a
strongly enhanced quenching rate. This will in turn result in a
decrease of the overall harvesting time $\tau$. For very large
biases, the overall harvesting time will eventually rise again, as a
result of the reduced relaxation rates $W_{ss'}$. All this is in
agreement with the behavior observed in Fig.~\ref{fig:quench2} for
$\Gamma \ll W_0$.

\begin{figure}[ht]
\centerline{\includegraphics[width=8cm]{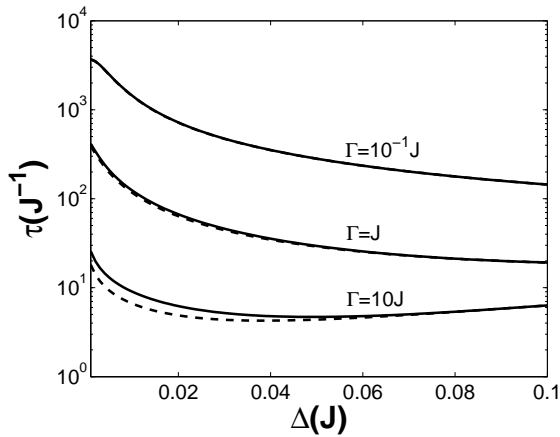}}
    \caption{Harvesting time $\tau$ [defined in Eq.~(\protect\ref{taucorrect})]
    versus the applied bias magnitude $\Delta$ calculated for a
    disorder-free chain of 50 sites. The solid lines correspond to the model
    where coherences are taken into account, while the dashed lines only account
    for populations and are identical to the results in
    Fig.~\protect\ref{fig:quench2}.}
    \label{fig:coh}
\end{figure}

The nonmonotonic $\Delta$-dependence of $\tau$ observed in
Fig.~\ref{fig:quench2} (reaching a minimal $\tau$ value at small
bias) occurs when the quenching and scattering amplitudes are
comparable, $\Gamma \sim W_0$. This nonmonotonic behavior is caused
by the competition between the bias-induced effects discussed above.
More specifically, for small but increasing values of $\Delta$, the
increase in the quenching rates $\Gamma_s$ for the exciton states
near the bottom of the band is the dominant effect. This explains
why $\tau$ initially decreases as a function of $\Delta$. A further
increase in $\Delta$ will shift the $\Gamma_s$-distribution to the
lower band edge, while the $P_s(0)$-distribution shifts to the upper
band edge, making the conditions for quenching less favorable. The
connection of these two distributions via the intra-band relaxation
accounts for the reduced harvesting efficiency beyond some optimum
bias value, due to the decrease of the relaxation rates for
increasing $\Delta$. Thus, we conclude that at $\Gamma \sim W_0$,
there exists an optimal magnitude of the bias that provides the most
efficient harvesting conditions.

It should be noted that for large values of the bias, when the
exciton states have become strongly localized, $\tau$ becomes
independent of the quenching amplitude, because the harvesting
process is limited in rate totally by the intraband relaxation. This
is confirmed by Fig.~\ref{fig:quench2}.

Figure~\ref{fig:coh} shows the corrections that a crude model (see
Sec.~\ref{Subsec: Validity}) for coherent transport yields, where
the time evolution of the coherences is assumed to be simply
exponential. The corrections are at most somewhere around $25$
percent, which makes sense as direct quenching only accounts for a
fraction of the total quenched population. Also, dephasing of the
coherences is provided through both the exciton-phonon coupling and
the trap, and coherent effects are therefore suppressed for all but
the fastest energy transfer processes. A full discussion of the
effects of coherent transport in unbiased chains should not be
limited to the secular treatment provided here, and will be provided
in a later paper.

\begin{figure}[ht]
\centerline{\includegraphics[width=8cm]{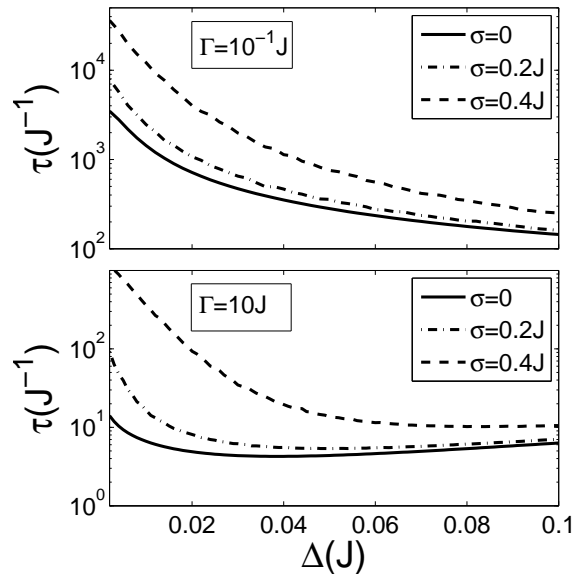}}
    \caption{Effect of uncorrelated on-site Gaussian disorder of
    strength $\sigma$ on the bias dependence of the harvesting time $\tau$
    calculated for two quenching amplitudes $\Gamma = 10^{-1}J$ (upper
    panel) and $\Gamma = 10J$ (lower panel).}
\label{fig:f5}
\end{figure}

\subsection{Disordered biased chain}
    \label{Subsec: Inhomogeneous}

In this section, we present a study of the bias-dependent harvesting
time $\tau$ in disordered chains. We considered on-site Gaussian
disorder (uncorrelated for different sites) with zero mean and
standard deviation $\sigma$, referred to as the disorder strength.
Figure~\ref{fig:f5} shows the results obtained for two quenching
amplitudes $\Gamma = 10^{-1} J$ and $\Gamma = 10J$, keeping $W_0 =
10 J$ and $N = 50$ in all simulations. For other values of the
scattering or quenching amplitude, similar results were obtained.
The general observation that can be deduced from Fig.~\ref{fig:f5}
is that the inclusion of disorder leads to a reduced efficiency of
the energy harvesting process. The effect is especially significant
for small $\Delta$. Another consequence of disorder is that, in the
case of $\Gamma \sim W_0$, the magnitude of the bias that is optimal
for harvesting shifts to larger values.

The explanation of these observations is straightforward. As we
already mentioned, the disorder results in Anderson localization of
all the states on segments of the chain. First of all, this
localization breaks the correlation that exists at low magnitudes of
the bias ($N \Delta \ll E_N - E_1 = 4.207 J$) between the
distributions $P_s(0)$ and $\Gamma_s$: the states that are mostly
excited initially and those that are quenched efficiently do not
overlap anymore. Secondly, the localized states that have a
significant amplitude at the trap are now not necessarily the lowest
energy states and can hardly be reached during the downward
intra-band relaxation. In addition, the localization slows down the
overall exciton relaxation, because of blocking of the diffusion
over the localized states at zero temperature.~\cite{Bednarz03} All
these factors result in a reduced harvesting efficiency at zero
bias. Higher bias strengths tend to (partially) restore the
correlation between the energy and location of an exciton state, in
particular the lowest-energy exciton states will tend to lie in the
vicinity of the trap. As a result, the states that are quenched most
strongly are more accessible through intraband relaxation. This is
why for small bias values $\Delta$, the harvesting time $\tau$
always decreases for increasing $\Delta$, independently of the
quenching amplitude $\Gamma$.

For larger bias magnitudes, such that $N\Delta > E_N - E_1 = 4.207
J$, the dynamic localization dominates over the disorder-induced
localization. As a consequence, in this region the effect of
introducing disorder is very limited, in full agreement with the
results presented in Fig.~\ref{fig:f5}. Note that, like the Bloch
localization discussed in the previous section and for similar
reasons, Anderson localization expands the range of validity of our
model.

\section{Summary and concluding remarks}
    \label{Sec: Summary}

We studied the efficiency of harvesting of excitation energy in
chains of strongly coupled dipolar units, in the presence of a
linear energy bias. The harvesting combines the excitation
transport following initial excitation at one end of the chain and
quenching of the excitation by a trap at the other end. The energy
gradient was set towards the trap. We considered the quenching
rate of a particular exciton state to be proportional to its
occupation probability at the position of the trap. Within this
model, the energy harvesting efficiency is governed by the
interplay of the direct quenching of exciton states due to their
overlap with the trap and the exciton relaxation to those states
which are quenched most efficiently. We found a complicated
scenario for the energy harvesting efficiency as a function of the
bias magnitude $\Delta$. Most importantly, we found that a bias
does not necessarily increase this efficiency.

The effect of a bias strongly depends on the ratio between the
amplitudes $\Gamma$ (efficiency of the trap) and $W_0$
(vibration-induced scattering). In the limit of $\Gamma \ll W_0$,
the initial population first scatters toward the bottom of the band,
where quenching (the slowest process) eventually takes place.
Introducing a bias now greatly reduces $\tau$, because of its above
mentioned tendency to shift the strongly quenched states to the
lower band edge. Suppression of the exciton's downward relaxation
upon increasing the bias magnitude now plays a less important role,
because the relaxation is not the rate limiting step in the process.

In the intermediate case of $\Gamma \sim W_0$, it depends on the
bias strength what is the rate limiting step in the harvesting
process. As a consequence, the bias dependence of the harvesting
time $\tau$ reveals a nonmonotonic behavior, in other words an
optimal magnitude of the bias exists at which excitons are harvested
most efficiently. As has been explained in Section \ref{Sec:
Population dynamics}, our model is not expected to be appropriate in
the limit of $\Gamma \gg W_0$, corresponding to what we expect is an
unphysically strong quenching effect, as it does not adequately
include coherent effects.

We have also investigated how random energy disorder, superimposed
on the overall bias, affects the above findings. We have found
that disorder leads to a decrease of the harvesting efficiency, in
particular for low bias magnitudes, and gives rise to a
nonmonotonic bias dependence of this efficiency, independently of
the ratio $\Gamma/W_0$. We have shown that these effects can be
understood from the disorder-induced (Anderson) localization of
the exciton states. The effects of disorder are smeared at higher
bias magnitudes, when the dynamic (Bloch) localization becomes
dominant.

To conclude, we note that the competition between the enhancement
of transport and quenching, which underlies the interesting
nonmonotonic behavior discussed in this paper, is a very common
theme in the physics of energy harvesting systems. Although the
model presented here is inspired by one-dimensional Frenkel
exciton systems, such as J-aggregates and conjugated polymers, it
is to be expected that similar effects occur in other physical
realizations, such as linear arrays of resonantly coupled quantum
dots.

\acknowledgments This work is supported by NanoNed, a national
nanotechnology programme coordinated by the Dutch Ministry of
Economic Affairs.

\end{document}